\documentclass{elsart}
\usepackage{graphicx}
\usepackage{latexsym}
\usepackage{amsmath}
\usepackage{amssymb}
\usepackage{longtable}
\usepackage{multirow}
\usepackage{lineno}
\begin{document}
\begin{frontmatter}
\title{First Study of Neutron Tagging with a Water Cherenkov Detector}
\author{H. Watanabe\thanksref{label1}\corauthref{cor1}}, 
\ead{h-watana@suketto.icrr.u-tokyo.ac.jp}
\author{H. Zhang\thanksref{label32}\corauthref{cor1}},
\ead{\\ \hspace{3.35cm}zhanghb02@mails.tsinghua.edu.cn}
\corauth[cor1]{Corresponding Author}
\author{K. Abe\thanksref{label1}},
\author{Y. Hayato\thanksref{label1}},
\author{T. Iida\thanksref{label1}},
\author{M. Ikeda\thanksref{label1}},
\author{J. Kameda\thanksref{label1}},
\author{K. Kobayashi\thanksref{label1}},
\author{Y. Koshio\thanksref{label1}},
\author{M. Miura\thanksref{label1}},
\author{S. Moriyama\thanksref{label1}\thanksref{label31}},
\author{M. Nakahata\thanksref{label1}\thanksref{label31}},
% ...
\\
(The Super-Kamiokande Collaboration)
\address[label1]{Kamioka Observatory, Institute for Cosmic Ray Research, The University of Tokyo, Hida, Gifu 506-1205, Japan}
% ...
\address[label31]{Institute for the Physics and Mathematics of the Universe (IPMU), The University of Tokyo, Kashiwa, Chiba 277-8568, Japan}
\address[label32]{Department of Engineering Physics, Tsinghua University, Beijing 100084, China}
% ...

\begin{abstract}
A first study of neutron tagging is conducted in Super--Kamiokande, a 50,000-ton water Cherenkov detector. The tagging efficiencies of thermal neutrons are evaluated in a 0.2 \% GdCl$_{3}$-water solution and pure water. They are determined to be, respectively, 66.7 \% for events above 3 MeV and 20 \% with corresponding background probabilities of 2 $\times$ 10$^{-4}$ and 3 $\times$ 10$^{-2}$. This newly developed technique may enable water Cherenkov detectors to identify $\bar \nu_{e}$'s geological or astrophysical sources as well as those produced by commercial reactors via the delayed coincidence scheme.
\end{abstract}

\end{frontmatter}

\section{Introduction}
\label{sec:intro}
\noindent
Light water Cherenkov (WC) detectors have been used for many years as inexpensive, effective detectors for neutrino interactions and nucleon decay searches. Examples include IMB, Kamiokande, and Super--Kamiokande (SK). While many important measurements have been made with these detectors (e.g. discovery of neutrino oscillations~\cite{osc}, discovery of neutrinos from stellar collapse~\cite{sn}, limits on nucleon decay~\cite{pdk}, determination of the solar neutrino oscillation parameters~\cite{solnu}), a major drawback of such light water detectors has been their inability to detect the absorption of thermal neutrons.

Neutrons liberated in water by the inverse beta reaction $\overline{\nu}_{e} + p \rightarrow e^+ + n$ (and other processes) are quickly thermalized. On average it takes about twenty collisions with the water's free protons over the course of $\sim$ 10 $\mu$s to bring a neutron emitted with a few MeV down to room temperature (0.025 eV).

Once thermalized, and after bouncing around for another 200 $\mu$s or so, the neutron is captured by a proton or oxygen nucleus in the water. The cross sections for these capture reactions are 0.33 barns and 0.19 millibarns, respectively, so to first approximation every thermal neutron is captured on a free proton via the reaction $n + p \rightarrow d + \gamma$.

The resulting gamma has an energy of 2.2 MeV and makes very little detectable light since the Compton-scattered electron is close to Cherenkov threshold. Hence, in traditional light water Cherenkov detectors (which have tended to have trigger thresholds around 5 MeV~\cite{nim}) these neutron captures are generally not recorded, and consequently there has been no way to tag anti-neutrino interactions on free protons. For comparison, detection of neutrons from neutral current interactions has already been demonstrated in a heavy water Cherenkov detector through mono-energetic 6.25 MeV gammas: $n + d \rightarrow t + \gamma\ ({\rm 6.25\ MeV})$~\cite{sno}.

In this paper we explore two independent approaches to extracting this neutron tagging information in SK, a 50,000 ton cylindrical water Cherenkov detector viewd by 11,129 20-inch diameter photo-multipliers. These studies collectively represent the first time that evidence of thermal neutron capture has ever been observed in a light water Cherenkov detector.

The first approach~\cite{gad} involved loading light water with a transparent, water soluble compound of gadolinium, gadolinium chloride (GdCl$_{3}$). Neutron capture on gadolinium yields a 7.9 MeV gamma cascade 80.5 \% of the time and a 8.5 MeV gamma cascade 19.3 \% of the time (hereafter, both cascades are referred to as 8 MeV gamma cascades). These relatively energetic gammas should be easily seen in SK and readily reconstructed as events with their own vertex position, time, and total energy, allowing event-by-event correlation with previous positron-like events' timing and position information. Natural gadolinium has a neutron absorption cross section of $\sim$ 49,700 barns as compared to 0.33 barns for free protons, so the amount of GdCl$_{3}$ needed for this method to be effective is quite small. To get over 90 \% of thermal neutrons to capture on gadolinium instead of hydrogen, 0.2 \% by mass of dissolved GdCl$_{3}$ ({\it i.e.}, about 0.1 \% Gd by mass) is sufficient.

The second approach~\cite{force} used 2.2 MeV gammas from $n + p \rightarrow d + \gamma$ reaction. Instead of adding anything to the pure water in the detector, it involved the introduction of special trigger electronics to force SK's existing data acquisition (DAQ) system to take 500 $\mu$s of data with no threshold requirement immediately after any primary events above a fixed, higher threshold were detected. While most triggers falling within this special gate cannot generally be reconstructed due to their very low energies, in this way some of the 2.2 MeV gammas released by neutron captures on hydrogen in pure water can be statistically identified as correlated in time with an energetic, primary event.  

In this paper, we present test results of both methods. Tagged neutrons were produced by using a bismuth germanate (BGO) scintillator cube with an Am/Be radioactive source embedded inside; the resulting scintillation light coupled with the emission of a neutron served to mimic inverse beta reactions occurring within the SK tank. Section \ref{sec:setup} is devoted to more details of the above experimental setup. Section \ref{sec:fog} introduces the newly-developed special trigger, or forced trigger, used in both approaches of data-taking. Data analysis and results with Gd are described in Section \ref{sec:gdana}, while Section \ref{sec:22ana} is dedicated to those employed with 2.2 MeV $\gamma$-rays. Finally, concluding remarks on this paper are established in Section \ref{sec:sum}.

\section{Experimental Setup}
\label{sec:setup}
\begin{figure}[htbp]
  \begin{center}
    \includegraphics[width=7.5cm]{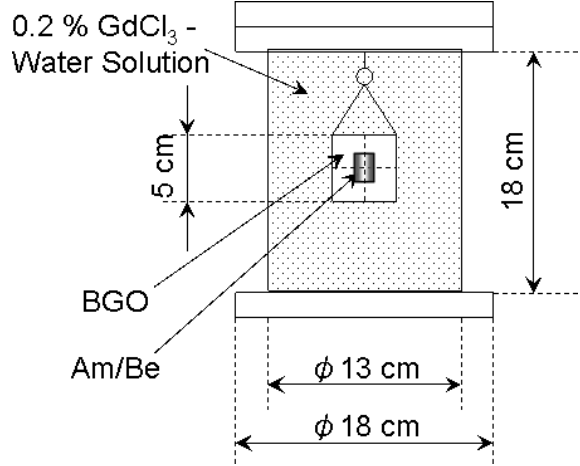}
    \caption{Configuration of the apparatus used for the study of 8 MeV gamma cascades from Gd. The acrylic cylindrical vessel was filled with 2.4 liters of 0.2 \% GdCl$_{3}$-water solution. The 5 cm cube of Am/Be-embedded BGO scintillator was installed in the middle of this cylinder. This BGO crystal with the Am/Be source was also used in the 2.2 MeV $\gamma$-ray search.}
    \label{fig:config}
  \end{center}
\end{figure}

\noindent
The apparatus used in both measurements was a 5 cm cube of BGO scintillator with an Am/Be radioactive source incorporated in its center. Both 4.43 MeV $\gamma$-rays and neutrons are released from the source concurrently via: $\alpha + {^{9}}{\rm Be} \rightarrow {^{12}}{\rm C}^{*}(4.43\ {\rm MeV deexcitation} \gamma) + n$ using $\alpha$'s emitted from $^{241}$Am. A reaction to the ground state of $^{12}$C also exists, which produces only a neutron. The source intensity used in the experiments is 97 $\mu$Ci of $^{241}$Am, which leads to 86.8 Hz of 4.43 MeV $\gamma$-rays and 76.4 Hz for the $^{12}$C ground state reaction. The scintillation light induced from these gammas were used to initiate prompt signals, while accompanying neutrons were treated as delayed signals. Usage of the BGO compensated for the disadvantage of low Cherenkov photon generation for the 4.43 MeV $\gamma$-rays in water. The scintillation light was typically observed as $\sim$ 1,000 photo-electrons, or equivalently, 1,000 PMT hits in SK. This high light output was capable of issuing the so-called ``high energy trigger,'' one of the trigger types used in the normal data-taking (see Section \ref{sec:fog}). Moreover, the 300 ns decay time constant of BGO scintillation light was long enough to discriminate it from Cherenkov events originating from either 8 MeV gamma cascades and 2.2 MeV gammas. An acrylic cylindrical vessel was used to study the detection of 8 MeV gamma cascades liberated from Gd. This vessel was filled with 2.4 liters of 0.2 \% GdCl$_{3}$-water solution together with the Am/Be-embedded BGO scintillator installed in the middle of the cylinder, as depicted in Figure \ref{fig:config}.

\section{Forced Trigger Module}
\label{sec:fog}
\begin{figure}
  \begin{center}
   \includegraphics[width=0.6\textwidth]{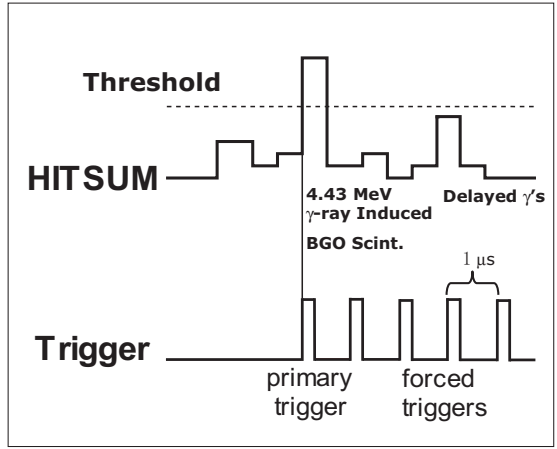}
    \caption{Time sequence of the triggers generated by the forced trigger module.}
    \label{fig:foglogic}
  \end{center}
\end{figure}

\noindent
In SK's normal data taking, there are three types of trigger signals derived from the total number of coincident PMT hits (HITSUM) in the inner detector. They are the ``Super Low Energy'' (SLE) trigger, the ``Low Energy'' (LE) trigger, and the ``High Energy'' (HE) trigger. The HITSUM, as illustrated in Figure \ref{fig:foglogic}, is the analog sum of all hit PMT signals within 200 ns. Once the HITSUM is above the designed threshold, a global trigger is issued and PMT data are digitized and stored in the internal buffers of the front-end electronics. PMT data are then read out and stored in the two 1-MByte memories of the Super Memory Partner (SMP), from which data will be transported to the online system. Since the lowest energy trigger thresold at SK (SLE, 4.6 MeV) is higher than 2.2 MeV, neutron signals in pure water cannot be observed. To measure neutron tagging efficiency, a standard nuclear instrument module (NIM) compatible with current SK electronics was developed by using a field programmable gate array (FPGA) chip and a micro-controller in order to provide selected triggers without threshold. This module can accept trigger signals from the SK DAQ system to issue a new trigger scheme as already shown in Figure \ref{fig:foglogic}. The forced trigger parameters such as forced trigger rate and the number of forced triggers can be programmed. During the experiments, 500 $\times$ 1 $\mu$s continuous forced triggers were generated after each primary trigger, whose repetition width was set on the basis of the electronics' dynamic range. Due to the limitations of the front-end electronics buffering capability and the two 1-MByte memories of the SMP, only about 128 triggers per primary were collected. The front-end buffering limitation caused a 70 $\mu$s gap between two bunches of forced trigger events. Since it is very likely that other trigger events might have occupied the front-end electronic buffer, the number of forced trigger events in the first bunch was usually less than 64. Due to the limitation of two buffer memories of the SMP and the speed of the reading process, all the triggers were vetoed after collecting about 128 forced trigger events. These limitations will be eliminated after the upgrade of SK electronics in September 2008, when 500 $\mu$s forced trigger data can be fully taken. As mentioned in Section \ref{sec:setup}, the 4.43 MeV $\gamma$-induced scintillation event rate was 86.8 Hz. To reduce dead time, we limited the primary trigger being used to generate the forced triggers to 4 Hz. 

\section{Study with Neutron Capture on Gadolinium}
\label{sec:gdana}
\noindent
In this study, the apparatus introduced in Figure \ref{fig:config} was deployed near the center of the SK detector. Described in the following subsections are data analysis and results with events collected by the forced trigger system.

\subsection{Distribution of Time Difference Between Prompt and Delayed Triggers}
\label{sec:gdana1}
\begin{figure}[htbp]
  \begin{center}
    \includegraphics[width=7.5cm]{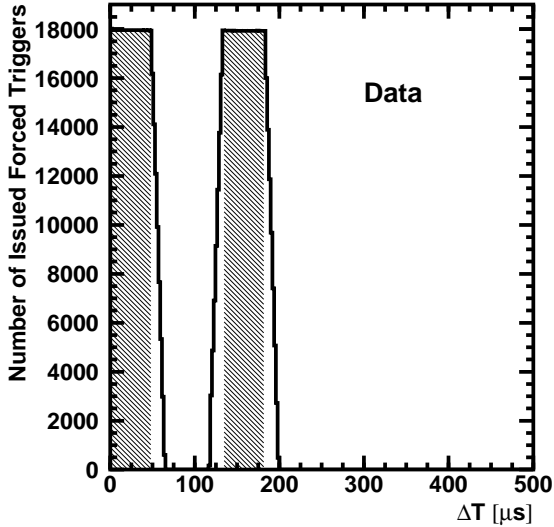}
    \caption{Distribution of $\Delta$T. $\Delta$T is defined as the time difference between the prompt (primary) and delayed (forced) triggers in terms of their time stamps. Trapezia indicate $\Delta$T with all issued delayed (forced) triggers. Superimposed shaded rectangles, on the other hand, are obtained with fully efficient activation of the data acquisition.}
    \label{fig:dt}
  \end{center}
\end{figure}

\noindent
The BGO scintillation light induced by the 4.43 MeV gammas produced the prompt (primary) triggers as described in Section \ref{sec:fog}. Due to the limited DAQ performance, available data existed within two 64 $\mu$s ranges measured from the prompt trigger with the presence of a 70 $\mu$s gap in between. A parameter $\Delta$T was characterized as a time difference between the prompt and delayed (forced) triggers in terms of their time stamps. This parameter was studied at the first stage of the data analysis to determine and extract the time ranges that resulted in full efficient data acquisition. Figure \ref{fig:dt} shows the result of $\Delta$T range with 100 \% efficiently activated data acquisition denoted by shaded rectangles: 1.2 $\mu$s $\sim$ 48.2 $\mu$s and 134.8 $\mu$s $\sim$ 181.8 $\mu$s. For comparison, the whole $\Delta$T distribution is the area under the solid black lines. Events collected in these time ranges were used for further stages of the data analysis.

\subsection{Vertex Distribution of Delayed Signals}
\label{sec:gdana2}
\begin{figure}[htbp]
  \begin{center}
    \includegraphics[width=7cm]{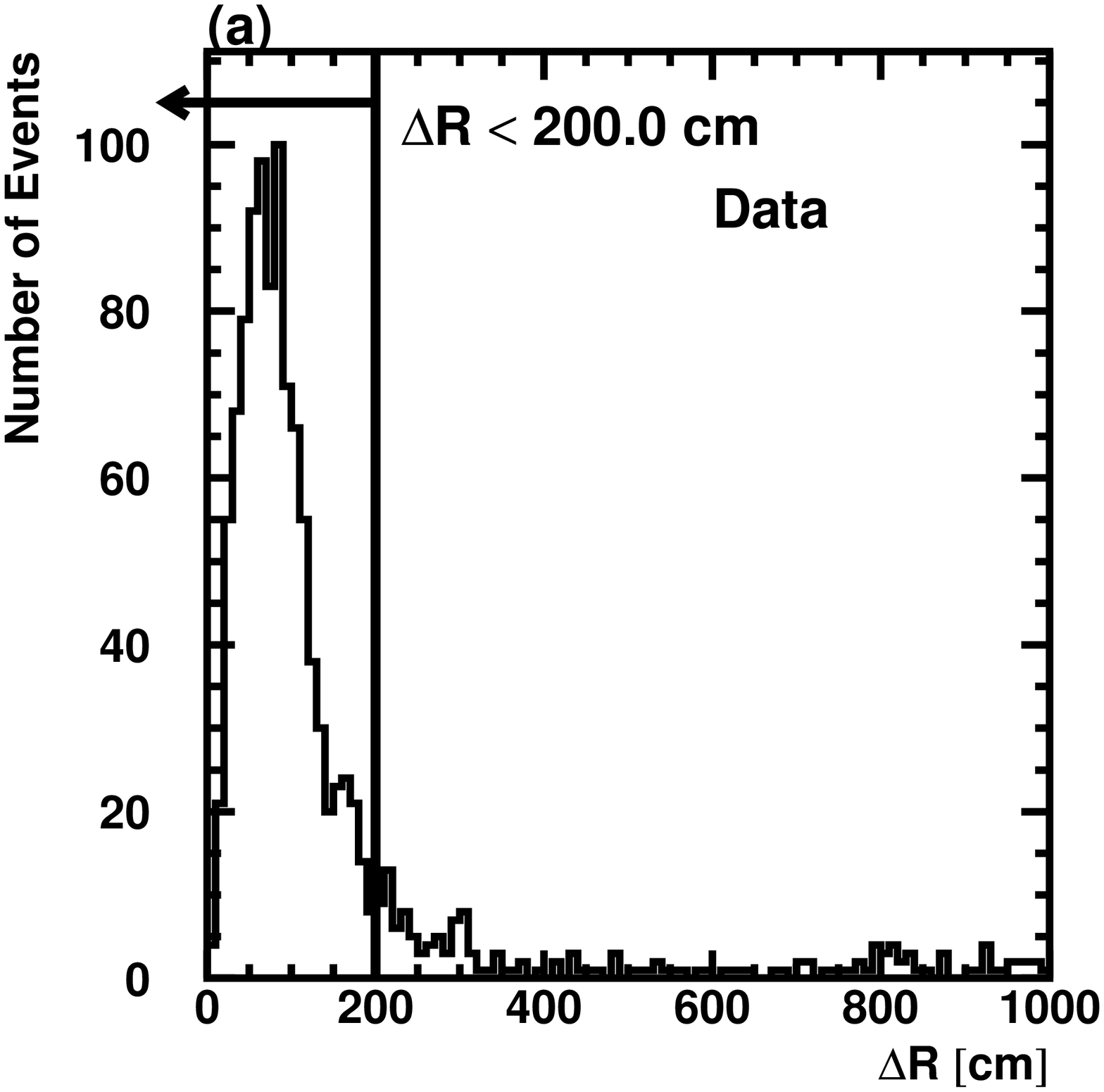}\includegraphics[width=7cm]{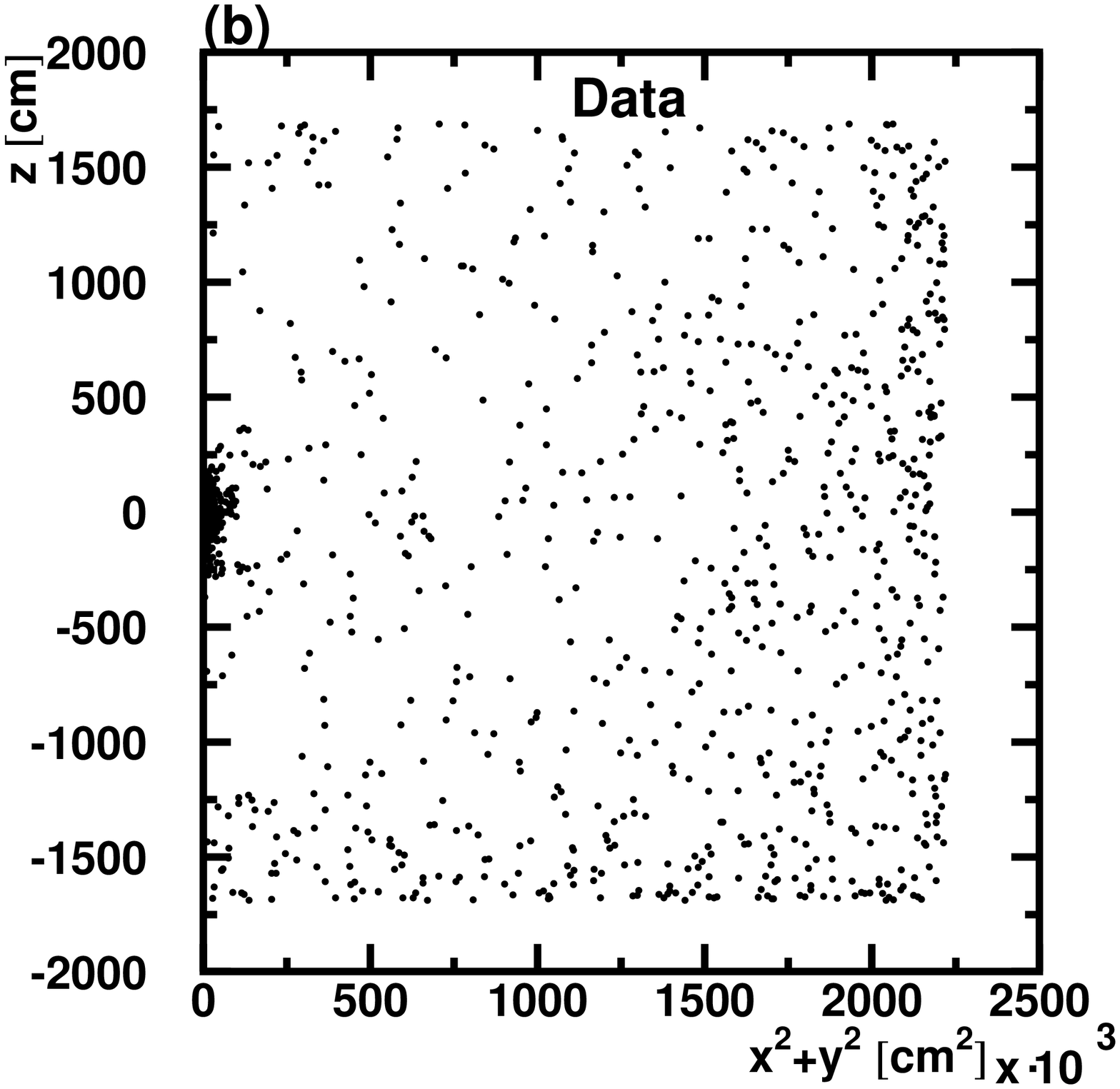}
    \caption{(a): Delayed signal $\Delta$R distribution of data. $\Delta$R in cm is the distance measured from the center of the GdCl$_{3}$ vessel ({\it i.e.}, Am/Be source position). (b): Correlation of $z$ [cm] versus $x^{2}+y^{2}$ [cm$^{2}$] of delayed events in data. These two plots illustrate the observation of 8 MeV gamma cascades with their vertices reconstructed around the Am/Be radioactive source position.}
    \label{fig:enedr}
  \end{center}
\end{figure}

\noindent
Events consistent with delayed 8 MeV gamma cascades were expected to be reconstructed around the Am/Be radioactive source. The histogram in Figure \ref{fig:enedr}(a) depicts the distribution of $\Delta$R = $\sqrt{x^{2} + y^{2} + z^{2}}$ in cm for delayed signal data, where $\Delta$R was the distance measured from the center of the GdCl$_{3}$ vessel ({\it i.e.}, Am/Be source position) whose coordinates were identified with ($x_{\rm src}, y_{\rm src}, z_{\rm src}$) = (35.3 cm, -70.7 cm, 0.0 cm). In SK, the $z$-axis was defined to run vertically, with its origin at the detector's center. Both $x$- and $y$-axes were perpendicular to each other and to the $z$-axis.  Therefore, the origin of our coordinate system (0, 0, 0) corresponded to the physical center of the SK detector. The majority of delayed signal events were observed in $\Delta$R $<$ 200.0 cm. A scatter plot of these is shown in Figure \ref{fig:enedr}(b) with respect to $z$ [cm] versus $x^{2}+y^{2}$ [cm$^{2}$], which clearly depicts the concentration of events the region $\Delta$R $<$ 200.0 cm. Detection of 8 MeV gamma cascades was therefore demonstrated through this vertex reconstruction in the vicinity of the Am/Be radioactive source. Event selections at this stage included the $\Delta$T ranges described in the previous section, the choice of reconstructed Cherenkov ring events, the energy region above 3 MeV, and the fiducial volume region which contained the innermost 22.5 kton cylindrical water region. The Cherenkov ring selection depends upon the pattern goodness of the fitted ring. This is evaluated via a determination of the goodness of direction reconstruction using the uniformity of hit PMTs about its presumed direction.

\subsection{Energy Spectrum and Thermal Neutron Capture Time for Delayed Events}
\label{sec:gdana3}
\begin{figure}[htbp]
  \begin{center}
    \includegraphics[width=7cm]{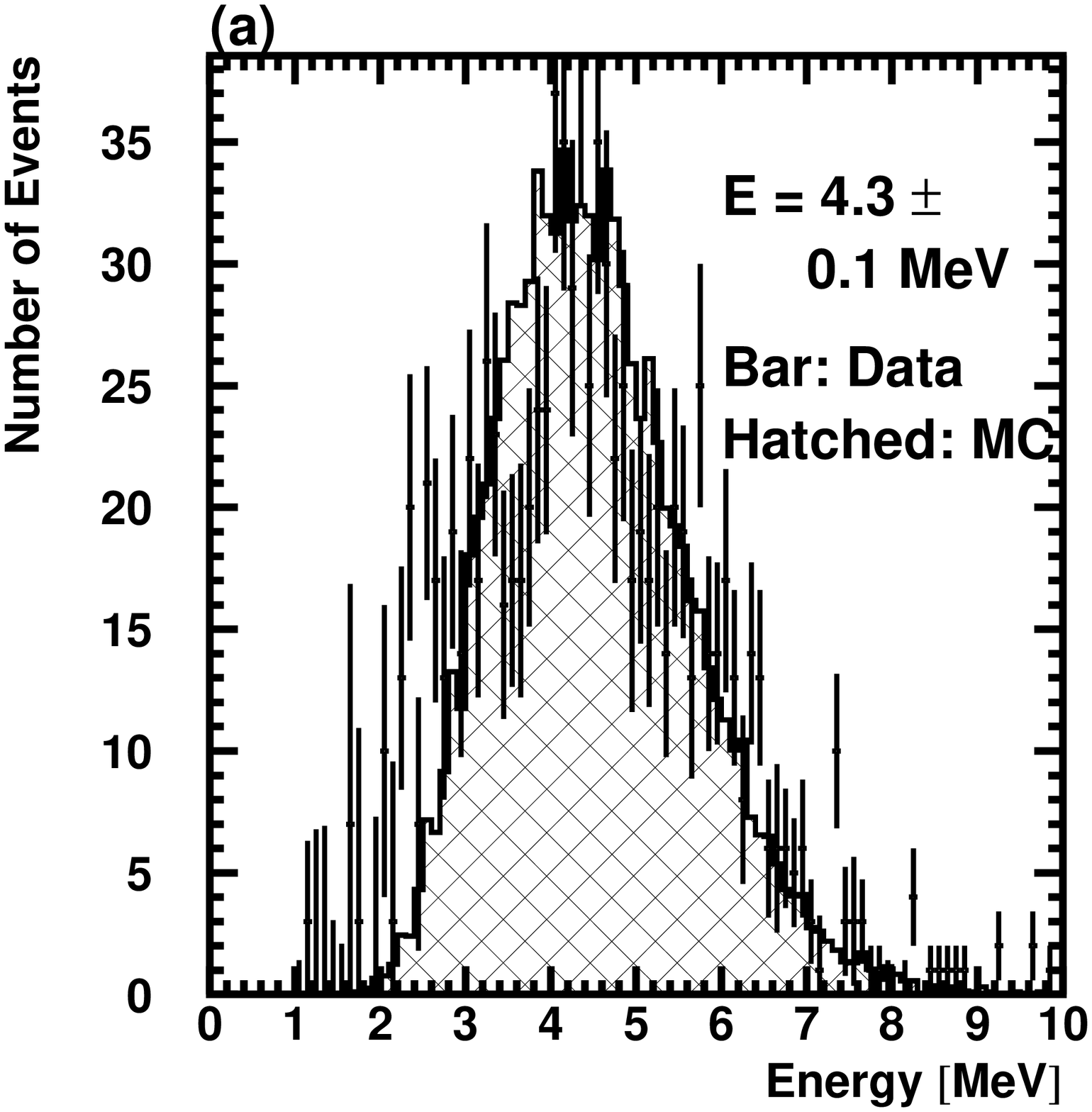}
    \includegraphics[width=7cm]{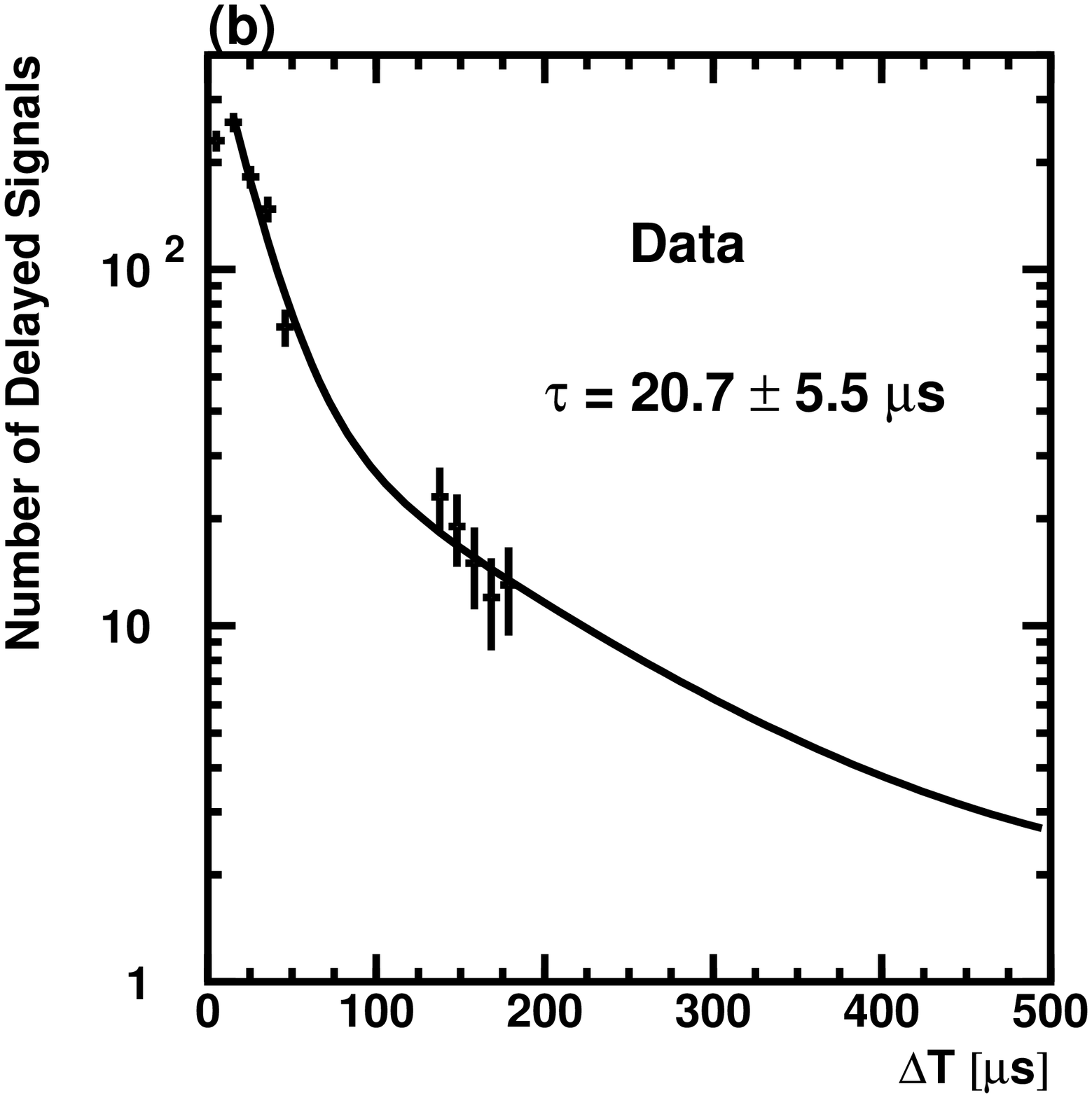}\includegraphics[width=7cm]{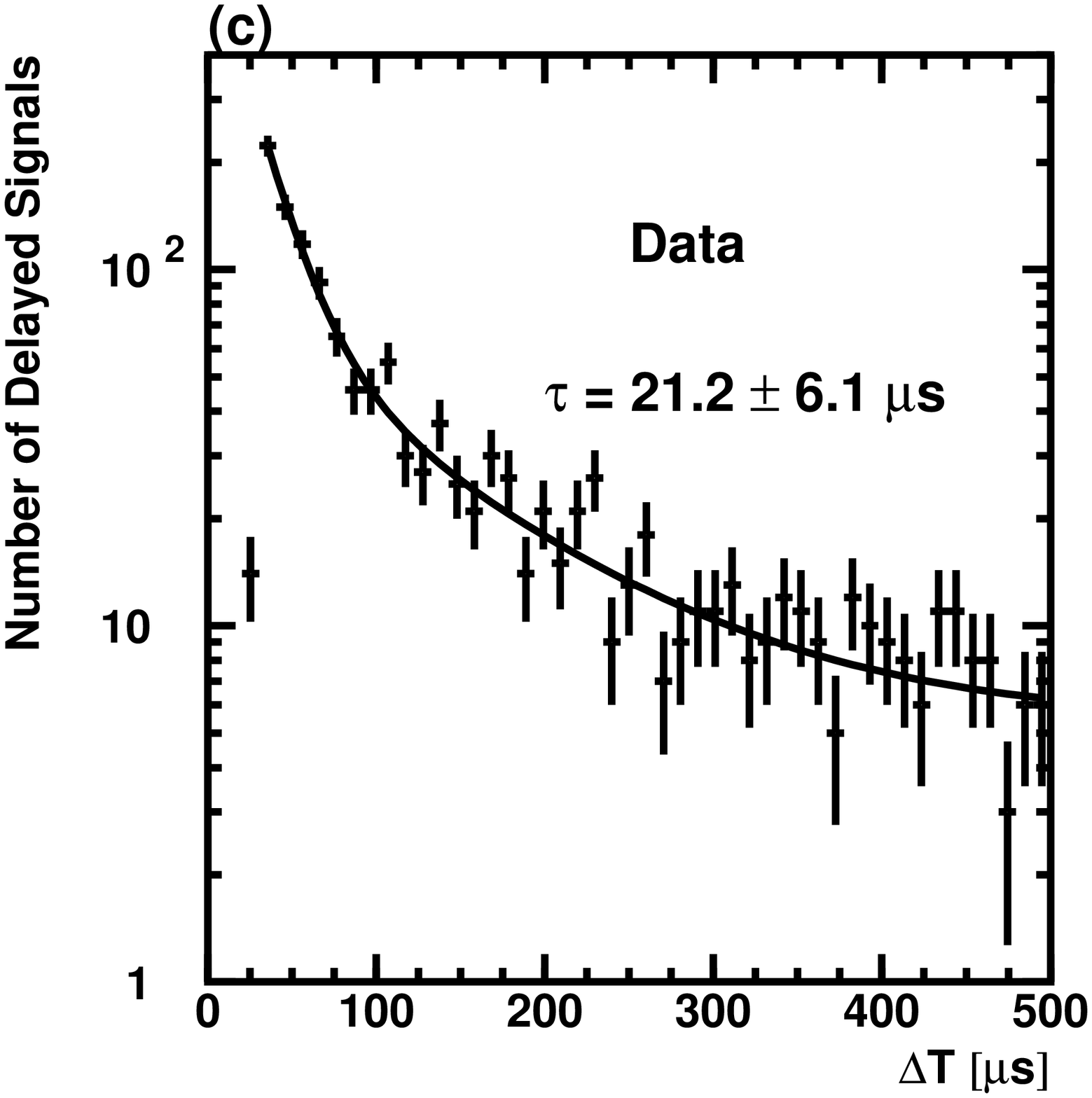}
    \caption{(a): Energy spectrum of data (bar) and MC normalized by the number of data events (hatched). These agree well each other; the Gaussian fitted mean energy value is 4.3 $\pm$ 0.1 MeV. (b): Thermal neutron capture time in data. (c): Capture time spectrum of data taken without using the forced trigger system. These fitted distributions with a sum of two exponentials both in (b) and (c) take into account long-duration thermal neutrons scattered back into the Gd area after having wandered in water. The capture time of the immediately captured thermal neutrons were determined from the first fitted component to be $\tau$ = 20.7 $\pm$ 5.5 $\mu$s and $\tau$ = 21.2 $\pm$ 6.1 $\mu$s, respectively. These corresponded to an MC result of $\tau$ = 20.3 $\pm$ 4.1 $\mu$s. The first observation of 8 MeV gamma cascades from gadolinium in a large water Cherenkov detector is thus demonstrated by these results.}
    \label{fig:reenetime}
  \end{center}
\end{figure}

\noindent
Figure \ref{fig:reenetime}(a) shows the energy spectrum for 8 MeV gamma cascades with the previously mentioned event selection criteria as well as the additional selection of $\Delta$R $<$ 200.0 cm applied. Data spectrum is displayed with bars. A Monte-Carlo (MC) simulation was also conducted, generating 8 MeV gamma cascades based upon GEANT4.7.1.p01 which in turn utilized the neutron database of G4NDL3.10. The generated 8 MeV gamma cascades within the 2.4 liter acrylic vessel were then incorporated into the standard SK MC simulation using GEANT3.21. The hatched histogram in Figure \ref{fig:reenetime}(a) is its resulting spectrum normalized with the number of data events for comparison. The data spectrum was obtained by subtracting the spectrum with a $\Delta$T range of 134.8 $\mu$s $\sim$ 181.8 $\mu$s from the data with a $\Delta$T range of 1.2 $\mu$s $\sim$ 48.2 $\mu$s in order to eliminate the contribution from non-8 MeV $\gamma$-ray-related events.

The data and MC spectra were found to be consistent with each other, and the mean energy value given by a Gaussian fit of the data was 4.3 $\pm$ 0.1 MeV. That this number was lower than 8 MeV was expected, given the various combinations of multiple $\gamma$-rays possible in the cascades. These gamma energies typically fall between 1 $\sim$ 2 MeV, with the lowest energy gammas unable to Compton scatter electrons above Cherenkov threshold; such gammas are invisible in SK. Figure \ref{fig:reenetime}(b) depicts the thermal neutron capture time of delayed events in data. The remaining events in the $\Delta$T range of 134.8 $\mu$s $\sim$ 181.8 $\mu$s resulted from thermal neutrons being scattered back into the Gd regions after having traveled out of the 2.4 liter cylinder. This phenomenon was studied with the capture time spectrum of data independently taken without using the forced trigger, as shown in Figure \ref{fig:reenetime}(c). Triggers of this data were issued by the standard SK trigger system which required at least 17 PMT hits within a 200 ns window as described in Section \ref{sec:fog}. The lowest level triggers, or ``super low energy'' triggers were vetoed for 30 $\mu$s right after the higher level ones (high energy triggers) were issued. Data in the range $\Delta$T $<$ 30 $\mu$s were unavailable with this data-taking method. Distributions both in Figure \ref{fig:reenetime}(b) and (c) were fitted with a sum of two exponentials, and the capture time of thermal neutrons immediately captured on Gd were derived from the first fitted component to be $\tau$ = 20.7 $\pm$ 5.5 $\mu$s and $\tau$ = 21.2 $\pm$ 6.1 $\mu$s, respectively. The result from MC was determined to be $\tau$ = 20.3 $\pm$ 4.1 $\mu$s. These consistent results of capture time constituted the first observation of 8 MeV gamma cascades from neutron capture on gadolinium in a large water Cherenkov detector.

\subsection{Estimation of Background Probability and Neutron Detection Efficiency}
\label{sec:gdana4}
\begin{figure}[htbp]
  \begin{center}
    \includegraphics[width=7cm]{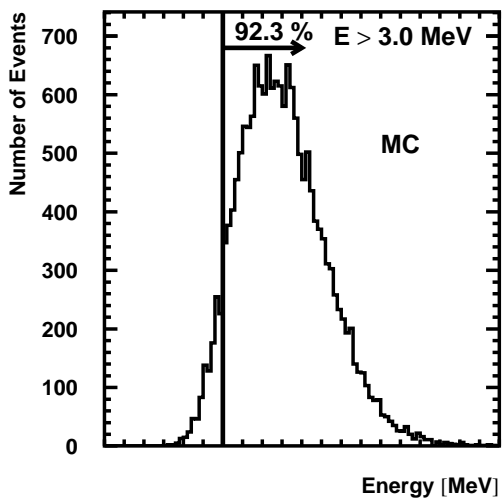}
    \caption{Energy spectrum for 8 MeV gamma cascades in MC. 92.3 \% detection efficiency is guaranteed above 3.0 MeV.}
    \label{fig:remcene}
  \end{center}
\end{figure}

\noindent
This section will discuss the probability of accidental coincident background events as well as the neutron detection efficiency. Understanding the background probability will be indispensable for the eventual future detection of relic supernova neutrinos, one of the principal physics interests in SK. The present singles background level above 10 MeV in SK fiducial volume is 8 events per day. The expected $\bar \nu_{e}$ event rate from relic supernovas is 7 events per year with neutron tagging. A reduction in background probability to $\sim$ 5 $\times$ 10$^{-4}$ times the current rate must therefore be achieved at this energy. In order to investigate the possibility of achieving this reduction, randomly triggered data in SK were taken. For this application, a 60 $\mu$s window was adopted together with all the other event selection criteria already established in the previous sections. A 60 $\mu$s window was sufficient to cover more than 90 \% of neutron capture events based upon the measured capture time. Table \ref{tbl:bgprob} summarizes the results of background probability at three different energy thresholds for delayed signals between 2.5 MeV and 3.5 MeV. The probability at 3.0 MeV of 2 $\times$ 10$^{-4}$ times the current rate leads to the conclusion that a 3.0 MeV threshold for delayed events will be satisfactory to accomplish $\bar \nu_{e}$ observations at 10 MeV. The neutron detection efficiency was evaluated including the effect of 8 MeV gamma cascade events extracted above 3.0 MeV. Figure \ref{fig:remcene} depicts an energy distribution for these events and 92.3 \% efficiency is maintained above this energy. Other components included were: 80.2 \% remaining after event selection, and 90 \% neutron capturing efficiency on Gd loaded at 0.1 \% by mass. The total detection efficiency of thermal neutrons on Gd was therefore 66.7 \%.

\begin{table}[htbp]
  \begin{center}
    \begin{tabular}{|c|c|} \hline
      Energy Threshold & Background Probability \\ \hline \hline
      2.5 MeV & $\sim$ 1 $\times$ 10$^{-3}$ \\ \hline
      3.0 MeV & $\sim$ 2 $\times$ 10$^{-4}$ \\ \hline
      3.5 MeV & $\sim$ 3 $\times$ 10$^{-5}$ \\ \hline
    \end{tabular}
    \caption{Summary of the background probability at three different energy thresholds for the delayed events. The study shows that 3.0 MeV for delayed signals will be satisfactory to enable ${\bar \nu_{e}}$ detection at a 10 MeV analysis threshold in SK.}
    \label{tbl:bgprob}
  \end{center}
\end{table}

\section{Identifying 2.2 MeV Gammas with a Special Trigger Logic}
\label{sec:22ana}
\noindent
Energy deposition of 2.2 MeV gammas in water is below the normal SK trigger threshold. To detect these extremely low energy gammas, a special trigger scheme, as already shown in Figure \ref{fig:foglogic}, was used to record possible neutron capture events. The BGO apparatus was placed inside SK at three different locations. Data taken under these conditions were referred to as source runs. For comparison, data were also taken in the same locations, but without the Am/Be source in the BGO apparatus. In such cases, the new trigger module generated 4 Hz of 500 $\times$ 1 $\mu$s forced triggers without requiring any primary trigger. Data taken under these conditions were referred to as background (BG) runs. In either run mode, two 64 event bunches are intended for analysis illustrated in Section \ref{sec:fog}.

In offline data analysis, it was required that a qualified primary event should have more than 1000 PMT hits and occur at least 1 ms after the previous event. The timing selection ensured that no neutrons from previous events would contaminate the sample. It was also required that forced trigger events should not have more than 200 PMT hits in order to eliminate other coincident but non-neutron events. Neutrons produced via inverse beta decays have a $\sim$ 50 cm free mean path and about a 200 $\mu$s lifetime in pure water. These features could be exploited. Monte Carlo study showed that knowing the capture location could significantly reduce background. In this study, we used the source position to subtract the time of flight for each timing measurement, correcting the timing spread up to 150 ns. This correction could restore all relevant hits to form a timing peak within 15 ns, which was significantly different from those from all the backgrounds. Selections on the number of PMT hits in a given time window and the time window width were optimized in order to give the best signal to noise ratio. 

\begin{figure}[htbp]
  \begin{center}
    \includegraphics[width=0.6\textwidth]{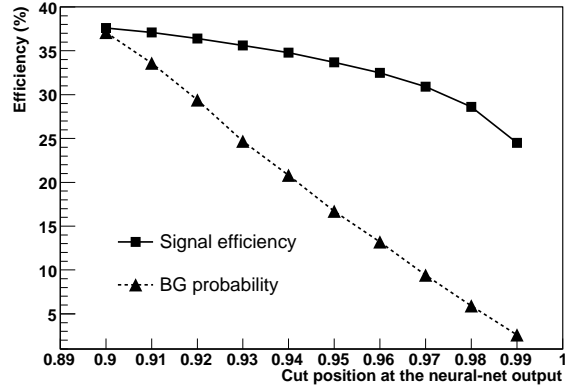}
    \caption{The neutron detection efficiency and the background probability as a function of the selection on the neural-net output value. Signal sample is from the Monte Carlo simulation for a 2.2 MeV gamma released at (35.3, -70.7, 1500.0) cm, while the background sample is from the background run.}
    \label{fig:nn_cut}
  \end{center}
\end{figure}

\begin{figure}
  \begin{center}
    \includegraphics[width=0.75\textwidth]{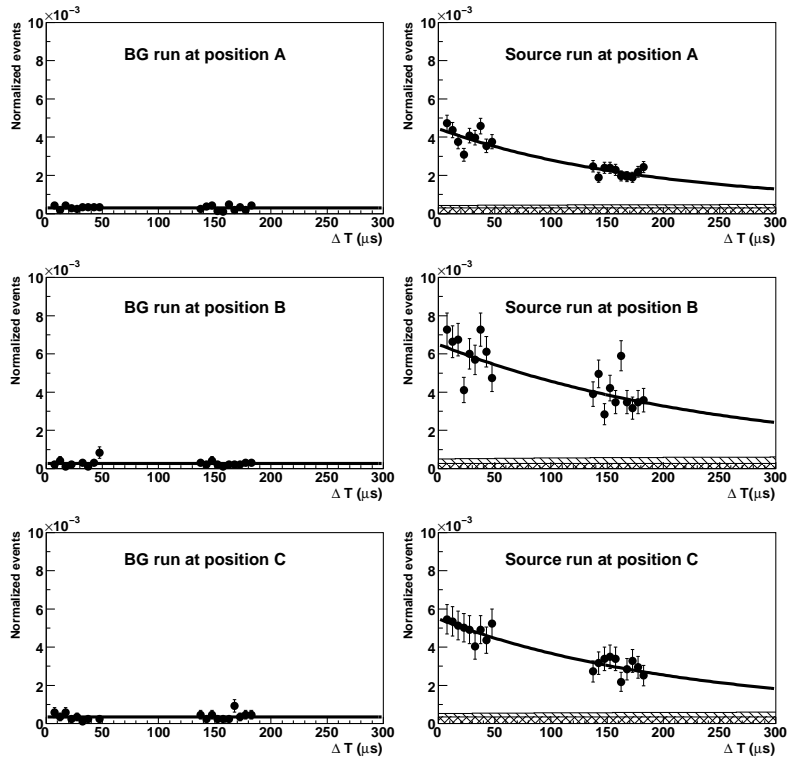}
    \caption{Normalized time distributions for both background runs and source runs at three different locations. The horizontal axis is the time difference between the neutron signal and the primary signal; the vertical axis is the entries normalized by total number of events. The position A, B and C are defined in Table \ref{lifetime}. Black dots are normalized measured values for background runs and source runs. The crossed histograms and hatched histograms represent source-related background and environmental background, respectively. Curves are fitted from histograms.}
    \label{fig:lifetime}
  \end{center}
\end{figure}

By comparing the PMT hit pattern of a signal with the known background, six discriminating variables were obtained to be used in further background reduction. They were the average opening angle of the hits, the angular uniformity of the hits, the number of hits within the 10 ns window ($N_{10}$), the fraction of the hits in the forward hemisphere ($N_{forward}/N_{10}$), the mean distance of the hits to their center of gravity, and the maximum distance among all the combinations of the hits. Some level of correlation was observed between these variables. A neural-net method implemented in ROOT~\cite{root} was then applied to optimize the selection criteria. The signal sample for neural-net training was from Monte Carlo simulation, while the background sample was from the background run, in which the background data were divided into a training sample and a test sample. Several neural-net architectures were tried and the two-hidden-layer network with six input neurons was selected for giving the best performance. The neural-net training procedure was conducted for each source location. The neutron detection efficiency and the background probability are shown in Figure \ref{fig:nn_cut} as a function of the selection on the neural-net output value. To select neutron signals from the source run data, all candidates were required to have a neural-net output value greater than 0.99. The $\Delta$T (Section \ref{sec:gdana1}) distribution is given in Figure \ref{fig:lifetime} for the three different locations. All source runs showed the expected exponential curves for the neutron capture lifetime in pure water. In contrast, the background runs only showed flat distributions due to the environmental background (Figure \ref{fig:lifetime}). The levels of contamination can be directly read from the vertical axes, since they were all normalized to the same number of primary triggers. 

To calculate the neutron detection efficiency, the source-related neutron background needed to be subtracted. At the event selection level, these neutrons looked exactly the same as the normal neutrons that were accompanied with BGO scintillation light. Since the timing is with respect to a wrong primary event, the distribution will be different from expected neutron lifetime distribution. To simplify the study, all the neutrons from the Am/Be source were classified into two categories: visible neutrons and invisible neutrons. The first category was for those produced via the process with an excited state of $^{12}$C$^{*}$ emitting a 4.43 MeV gamma which became the detected particle. The second category included neutrons whose associated 4.43 MeV gamma was not detected as well as those which had no associated gamma. Signal neutrons fall into the first category, but they only account for a small fraction since the primary trigger rate was prescaled down to 4 Hz. Most of visible neutrons in the first category are background neutrons for this study and all neutrons of the second category are background neutrons. Using the estimated detection efficiency in the BGO scintillator apparatus and the estimated production rates for both visible neutrons and invisible neutrons, the source-related neutron background rate at a given forced trigger time $t$ relative to the primary event was calculated to be:

\begin{equation}
R_{n}(t)=R^{inv}_{n}+R^{vis}_{n}\left[1-\exp\left({\frac{t}{\tau}}\right)\right],
\end{equation}

where the rates of $R^{inv}_{n}$ and $R^{vis}_{n}$ were estimated to be 76.4 Hz and 86.8 Hz, respectively. The $\tau$ was neutron capture lifetime in pure water. In the fit to the timing distributions without the environmental background, the following probability $P(t)$ was used:

\begin{equation}
P(t)=\alpha\cdot \exp\left(-\frac{t}{\tau}\right)+\alpha\cdot\tau\cdot R_{n}(t)
\cdot 10^{-6},
\end{equation}  

where the $10^{-6}$ was a factor to convert the neutron rate to $\mu$s units, and $\alpha$ was a fit parameter related to the neutron detection efficiency: 

\begin{equation}
\epsilon=\frac{\alpha\cdot\tau}{\mbox{time~bin~width}}.
\end{equation}

where the time bin width is 5 $\mu$s in this analysis. The results of the fits are listed in Table \ref{lifetime}. All the neutron capture lifetime values were consistent with the expected $\sim$ 200 $\mu$s. The measured neutron capture lifetime at SK using this method was 

\begin{equation}
\tau = (207 \pm 17_{stat.})\mu s.
\end{equation}

The neutron detection efficiencies at three different locations were derived using a fixed 200 $\mu$s lifetime as shown in Table \ref{lifetime}. Also shown are the estimated efficiencies from Monte Carlo simulations, the environmental background probabilities and the goodness of the fits. The measured neutron efficiencies are in agreement with those expected from Monte Carlo simulation. The efficiency is observed to increase when the capture location is close to the wall of the inner tank, since the small amount of light produced by the 2.2 MeV gammas suffers less attenuation effects. The position-independent background probability is considered to be approximately 3 $\times$ 10$^{-2}$. Backgrounds are originated from $\gamma$'s from the rock surrounding the detector, radioactive decay in the PMT glass, and radon contamination in the water.

\begin{table}
 \begin{center}
 \caption{Fitted neutron lifetimes and detection efficiencies at various $(x, y, z)$ locations: A = (35.3, -70.7, 0.0) cm; B = (35.3, -70.7, 1500.0) cm; and C = (35.3, -1201.9, 0.0) cm. A fixed 200 $\mu$s neutron capture lifetime is used to derive the neutron detection efficiencies. The first uncertainties are statistical, the second and the third uncertainties correspond to 20 $\mu$s lifetime uncertainty and 30 Hz neutron emission rate uncertainty. For comparison, the estimated efficiencies from Monte Carlo simulations, the environmental background probabilities and the goodness of the fits are given in the table.}
 \begin{tabular}{c|c|c|c|c|c} \hline \hline
 Location  & $\tau$ ($\mu$s)    & $\epsilon$(\%) 
 & M.C. $\epsilon$(\%)& Bkg prob.(\%)          & $\chi^2$/ndf \\
\hline
 \multirow{2}{*}{A} 
 & $192.0\pm16.2$ & $14.4\pm1.4$ & -            & -            &$22.3/17$ \\
 & fixed          & $14.7\pm0.4^{+0.5+0.2}_{-0.6}$ 
 & $13.1\pm0.2$ & $3.0\pm0.3$  & $22.6/18$ \\
 \hline
 \multirow{2}{*}{B} 
 & $256.4\pm33.6$ & $26.9\pm4.0$ & -            & -            & $29.6/17$ \\
 & fixed          & $24.4\pm0.9^{+1.0+0.2}_{-1.1}$ 
 & $24.5\pm0.2$ & $2.6\pm0.4$  & $33.4/18$ \\
 \hline
 \multirow{2}{*}{C} 
 & $221.5\pm29.9$ & $20.0\pm3.1$ & -            & -            & $7.6/17$ \\
 & fixed          & $19.2\pm0.8^{+0.7+0.1}_{-0.9}$ 
 & $17.6\pm0.2$ &  $3.0\pm0.3$ & $8.1/18$ \\
 \hline
 \end{tabular}
 \label{lifetime}
 \end{center}
\end{table}

\section{Summary}
\label{sec:sum}
\noindent
A first observation of neutrons has been established in a large light water Cherenkov detector using two different techniques. One of the techniques was the use of 2.4 liters of 0.2 \% GdCl$_{3}$-water solution which yielded 8 MeV $\gamma$-ray cascades from neutron captures on gadolinium. A study using this technique achieved 66.7 \% neutron tagging efficiency with a 3 MeV energy threshold for delayed events. Moreover, a background reduction level of 2 $\times$ 10$^{-4}$ at a 10 MeV prompt event analysis threshold for $\bar \nu_{e}$'s was realized. 

The other technique, a 2.2 MeV $\gamma$-ray search for neutrons captured on free protons, was also performed. This study revealed position-dependent detection efficiencies of neutrons ranging from 13.1 \% to 24.5 \% with the aid of a forced trigger system. These values indicated an approximate neutron detection efficiency of 20 \% assuming 2.2 MeV gammas are uniformly produced in the SK detector.
The background reduction level with this study is found to be 3 $\times$ 10$^{-2}$. 

Results from these research and development activities have verified the future ability to select $\bar \nu_{e}$'s at 10 MeV or even lower energy thresholds in large water Cherenkov detectors.

\section*{Acknowledgements}
\label{sec:acknow}
\noindent
The authors gratefully acknowledge the cooperation of the Kamioka Mining and Smelting Company. Super-Kamiokande has been built and operated with funds provided by the Japanese Ministry of Education, Culture, Sports, Science and Technology, the U.S. Department of Energy, and the U.S. National Science Foundation. This work has been partially supported by the Korean Research Foundation (BK21), the Korean Ministry of Science and Technology, the National Science Foundation of China, and Grant-in-Aid for Scientific Research in Japan.

\end{document}